\documentclass[prl,aps,twocolumn,nofootinbib,showpacs,groupedaddress]{revtex4-2}
\usepackage[usenames,dvipsnames]{color}
\usepackage[colorlinks=true,linkcolor=Violet,citecolor=Violet,urlcolor=Violet]{hyperref}
\usepackage{comment,graphicx,physics,booktabs}
\usepackage{bbm}
\usepackage{bm}
\usepackage{amsmath}
\usepackage{amssymb}
\usepackage{mathptmx}
\usepackage[version=3]{mhchem}

\renewcommand\vec{\boldsymbol}

\begin{document}
\title{Resonant Zener Interferometry in van der Waals Heterostructures}
\author{Nisarga Paul}
\email{npaul@caltech.edu}
\affiliation{Department of Physics and Institute for Quantum Information and Matter,
California Institute of Technology, Pasadena, California 91125, USA}
\author{Gil Refael}
\email{refael@caltech.edu}
\affiliation{Department of Physics and Institute for Quantum Information and Matter,
California Institute of Technology, Pasadena, California 91125, USA}

\date{\today}
\begin{abstract}
We demonstrate the presence of quantum interference effects in van der Waals heterostructures subject to in-plane electric fields. The in-plane field $F$ accelerates carriers through a hybridized band edge, and interlayer Zener tunneling occurs by distinct pathways, resulting in a solid-state quantum interferometer with imprints in transport observables. For parabolic-band bilayers, we identify two characteristic signatures which are observable in lateral conductance: Landau–Zener–St\"uckelberg oscillations in the band-overlap regime periodic in $1/F$ at small fields, resembling electric-field induced quantum oscillations, and a pronounced resonance at $F\propto T_0^{3/2}$ set by the interlayer hybridization $T_0$. These features provide a directly accessible probe of coherent interferometric dynamics in van der Waals heterostructures, and could be harnessed for more precise engineering and characterization.
\end{abstract}

\maketitle

\paragraph*{Introduction.} Charge creation from quantum tunneling is one of the most fundamental nonperturbative processes arising from light-matter coupling. Quantum electrodynamics, for instance, allows the creation of electron-positron pairs from the vacuum in a strong electric field by the Schwinger effect~\cite{sauter1931verhalten,Schwinger1951Jun}. In solids, the analogous phenomenon is Zener tunneling, in which an electric field drives carriers across a bandgap, ultimately causing dielectric breakdown~\cite{zener1934theory}, and phenomena directly parallel to the Schwinger effect have been observed, for instance, in graphene~\cite{schmitt2023mesoscopic}.\par

Zener tunneling is conventionally held to increase monotonically with the electric field, eventually leading to a breakdown of semiclassical electron dynamics. This view, however, neglects coherent phases accumulated by the tunneling electrons, which can play an important role~\cite{Nandkishore2011Aug,johri2013common,PhysRevB.74.041403}. In this work, we demonstrate that in two-dimensional semiconductor heterostructures~\cite{geim2013van,novoselov20162d,castellanos2022van}, phase coherence results in a solid-state quantum interferometer tunable by electric field. In particular, interlayer charge transmission becomes strongly non-monotonic, exhibiting a tunneling-dependent resonance as well as band-dispersion-dependent $1/F$ oscillations as a function of in-plane electric field $F$. These phenomena are experimentally accessible and depend distinctively on device characteristics, and hence we propose that this \textit{resonant Zener interferometry} can be harnessed for device characterization, calibration, and controllable interlayer electron-hole pair population.
\par
\begin{figure}
    \centering
\includegraphics[width=\linewidth]{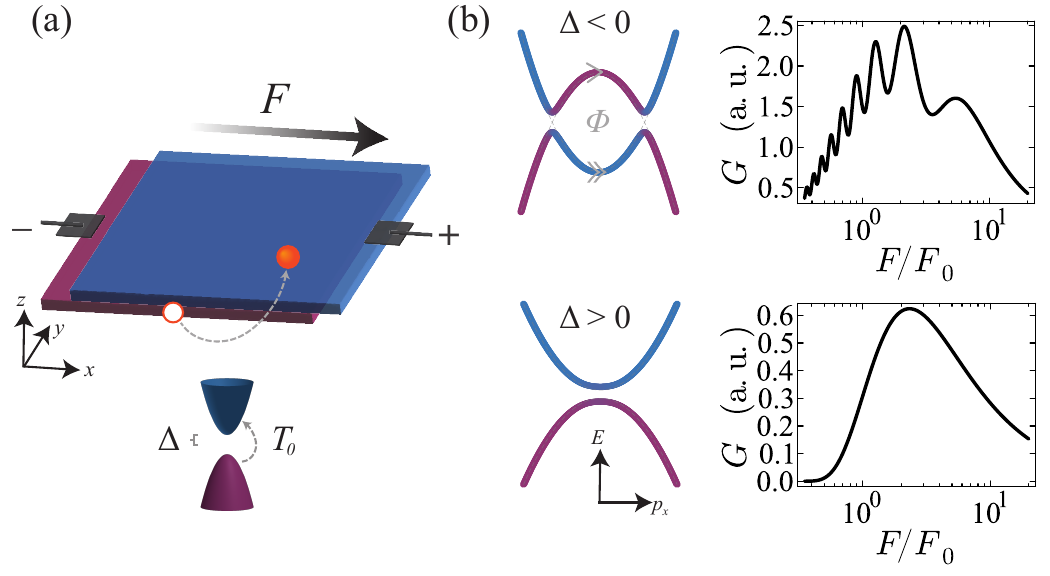}
\caption{(a) Schematic of a semiconductor van der Waals heterobilayer in an in-plane electric field $F$. Interlayer bias $\Delta$ and tunneling $T_0$ control field-driven interlayer charge transfer, measured by two-terminal conductance $G$. (b) For $\Delta <0$, $G$ exhibits interference-induced oscillations with period $\sim F^{2}/m^{\ast 1/2}|\Delta|^{3/2}$ at small fields (\textit{i.e.} periodic in $1/F$), with $m^*$ the effective mass. For all $\Delta$, $G$ shows a resonant peak at a characteristic field $F_0\sim m^{\ast 1/2}T_0^{3/2}$. These features realize a tunable solid-state interferometer whose signatures can probe $T_0$ and other device characteristics. }
    \label{fig:fig1}
\end{figure}
We focus on type-II aligned semiconductor devices, such as certain transition metal dichalcogenide (TMD) heterobilayers, where electron-hole creation is of significant interest due to the potential for long-lived interlayer excitons and resultant correlated phases~\cite{rivera2018interlayer,PhysRevLett.121.026402,devakul2021magic,mak2022semiconductor,regan2022emerging,qi2023thermodynamic}. We consider a heterobilayer (Fig.~\ref{fig:fig1}) subject to an in-plane electric field $F$ with lateral contacts. In this context, we uncover two concurrent phenomena which arise from subtle interference effects and produce robust, resolvable features in the conductance. \textit{(i)} When bands overlap ($\Delta<0$), there are oscillations periodic in $m^{\ast 1/2}|\Delta|^{3/2}/F$ at small fields where $m^*$ is effective mass, reminiscent of magnetic-field quantum oscillations but periodic in inverse \textit{electric} field. \textit{(ii)} For any bandgap $\Delta$, there is a resonant electric field $F_0\sim T_0^{3/2}$, where $T_0$ is the interlayer tunneling strength, at which conductance is maximized with simple functional form. These provide clear, novel diagnostics for model-specific parameters such as $m^*$, $\Delta$, and $T_0$, which can be otherwise difficult to measure; in particular, this may provide the first experimental method for directly measuring $T_0$, which is typically estimated from \textit{ab initio} calculations at present. Further features can also discriminate between $s$- and $p$-wave interlayer tunneling. The in-plane fields required for observation are within an achievable range ($10^5$-$10^7$~V/m) due to the meV-scale van der Waals hybridization energies. Overall, resonant Zener interferometry comprises qualitatively new quantum oscillations and nonlinear conductance signatures arising from quantum interference effects uniquely accessible in van der Waals devices. As such, it offers an avenue for enhanced control and characterization of these devices.
\par

The physical origin of the interferometric oscillations and resonance can be understood from simple two-level scattering dynamics. A uniform in-plane field sweeps valence band electrons through the
interlayer hybridized band edge, realizing coherent two-level scattering controlled by a sweep
rate $F$, the tunneling strength $T_0$, and interlayer bias $\Delta$. For $\Delta<0$ the electrons sweep through two avoided crossings (Fig.~\ref{fig:fig1}), and the resulting scattering probability oscillates due to the phenomenon of Landau-Zener-St\"uckelberg interferometry~\cite{shevchenko2010landau}. For $\Delta>0$, there are no real avoided crossings, but a nonadiabatic transition probability still exists due to \textit{complex}-time tunneling pathways, which interfere to produce a maximum at $F_0\sim T_0^{3/2}$. These
features are imprinted in the interlayer conductance, which is readily observable in van der Waals heterostructures under experimentally accessible conditions.

\paragraph*{Model.}
We consider an electron--hole bilayer in a uniform in-plane field $F\hat x$, described in the temporal gauge by
$H(t)=H_0(\vec p+e\vec A(t))$ with $\vec A(t)=-Ft\,\hat x$ ($e>0$).
In the layer (equivalently, conduction/valence band) basis,
\begin{equation}\label{eq:H0main}
    H_0(\vec p)=
    \begin{pmatrix}
        p^2/2m_{\mathrm c}+\Delta & T_{\vec p}\\
        T_{\vec p}^* & -p^2/2m_{\mathrm v}-\Delta
    \end{pmatrix},
\end{equation}
where $\vec p=(p_x,p_y)$, $m_{\mathrm c,\mathrm v}$ are effective masses, $\Delta$ is the interlayer bias, and
$T_{\vec p}$ is the interlayer tunneling.
In TMD heterobilayers $T_{\vec p}$ generically contains $s$- and chiral $p$-wave components~\cite{Rivera2018Nov,Zeng2023Nov}, which we parameterize as
\begin{equation}\label{eq:tunneling}
    T_{\vec p}=T_0+T_1(p_x+i\nu p_y),
\end{equation}
with spin/valley index $\nu=\pm1$. Henceforth we fix $\nu =1$, as later results apply independently of valley.
\par 
For clarity, we set $m_{\mathrm c}=m_{\mathrm v}=m^*$, though our conclusions are qualitatively unchanged for $m_{\mathrm c}\neq m_{\mathrm v}$. We introduce the Zener length and energy scales
\begin{equation}
    \ell_{\mathrm Z} =\left(\hbar^2/2m^*eF\right)^{1/3}= k_\mathrm{Z}^{-1},\quad
    E_{\mathrm Z} =eF\ell_{\mathrm{Z}}= \hbar t_{\mathrm{Z}}^{-1}.
\end{equation}
Because the field points along $\hat x$, $p_y$ is conserved. The time-dependent Schr\"odinger equation
$i\hbar \partial_t \psi_{\vec p}(t)=H_0(\vec p-e\vec A(t))\psi_{\vec p}(t)$ reduces to a simple two-level scattering problem for each $p_y$, in which an incoming valence band electron is driven by the field through the region of minimal gap. Each $p_y$ contributes a nonadiabatic interlayer transition probability, \textit{i.e.} the probability for a carrier injected in the valence layer to emerge in the conduction layer.

\paragraph*{Interlayer two-terminal conductance.}
We consider transport between lateral contacts, as shown in Fig.~\ref{fig:fig1}, which we calculate using the Landauer-B\"uttiker (LB) formula~\cite{landauer1957spatial}. LB applies to scattering between leads at different chemical potentials, while here we are considering an in-plane field, so it is not obvious that it applies. Nevertheless, the LB formula carries over to this case, as we show in the Supplemental Material~\cite{supp}. The LB conductance is:
\begin{equation}\label{eq:GmodeSum_general}
    G=\frac{ge^2}{h}\,\frac{L_y}{2\pi \hbar}\int \mathrm{d}p_y\;\mathcal{T}_{p_y},
\end{equation}
where $g$ is the spin/valley degeneracy, $L_y$ is the sample width, and $\mathcal{T}_{p_y}$ is the transmission probability for channel $p_y$. $\mathcal{T}_{p_y}$ is the probability that an electron at momentum $p_y$ injected into the valence layer emerges in the conduction layer.\par

\begin{figure}
    \centering
\includegraphics[width=\linewidth]{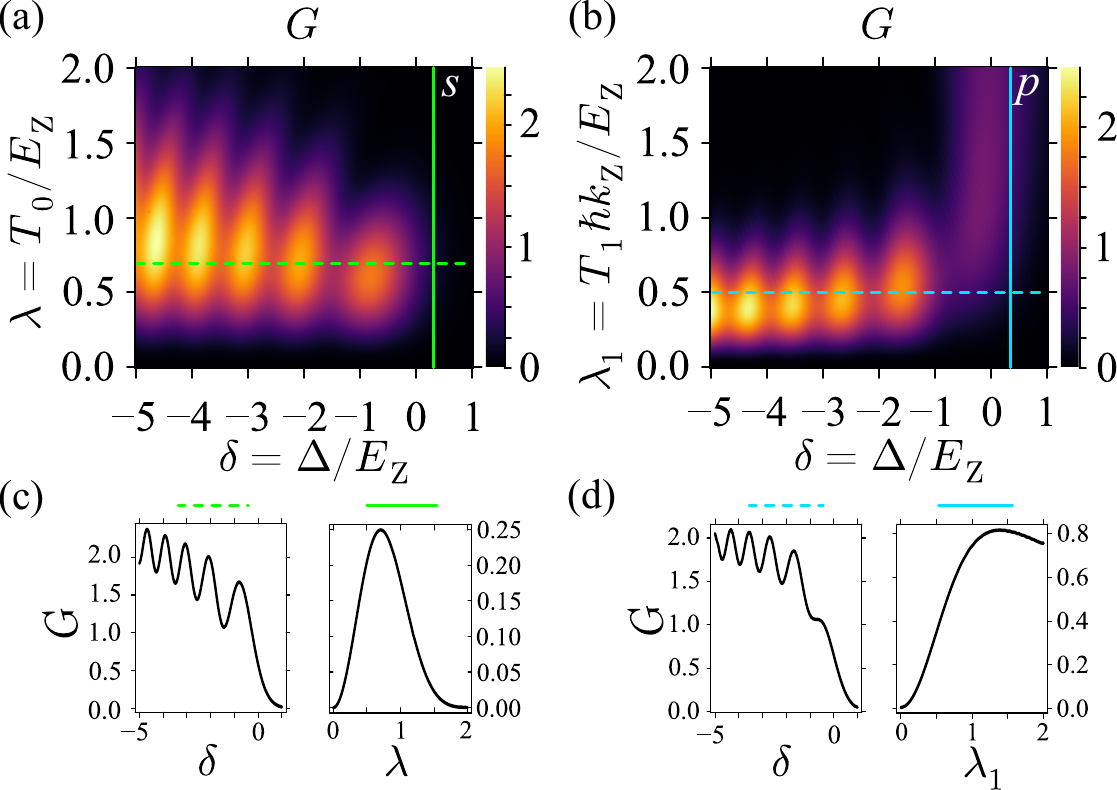}  
\caption{Lateral conductance $G$ (in units $ge^2 L_y k_Z/h$) for an electron-hole bilayer in an in-plane electric field $F$. (a) For $s$-wave tunneling $T_0/E_Z = \lambda$ and interlayer bias $\Delta/E_Z = \delta$, where $E_Z = (\hbar^2e^2 F^2/2m^* )^{1/3}$ is the energy scale set by $F$. (b) For $p$-wave tunneling $T_1\hbar k_Z/E_Z = \lambda_1$, where $k_Z = eF/E_Z$. (c) Line cuts for $s$-wave at $\lambda = 0.7$ and $\delta = 0.3$, showing Landau--Zener--St\"uckelberg oscillations and a resonant peak in conductance, respectively. (d) Line cuts for $p$-wave at $\lambda_1 = 0.5$ and $\delta = 0.3$, respectively, showing similar behavior.}
    \label{fig:conductance}
\end{figure}

For $s$-wave tunneling $T_{\vec p}=T_0>0$, the problem is characterized by the dimensionless parameters
\begin{equation}
    \kappa_y = p_y/\hbar k_\mathrm{Z}, \quad \delta=\Delta/E_{\mathrm Z},\quad
    \lambda=T_0/E_{\mathrm Z},
\end{equation}
and Eq.~\eqref{eq:GmodeSum_general} can be written as
\begin{equation}\label{eq:GmodeSum}
\begin{aligned}
    G =\frac{ge^2}{h}\,\frac{L_y k_{\mathrm Z}}{2\pi}\int \mathrm{d}\kappa_y\;
    P\!\bigl(\delta+\kappa_y^2,\lambda\bigr).
\end{aligned}
\end{equation}
Here, $P(\delta,\lambda)$ (to be defined in Eq.~\eqref{eq:Pdeltalambda}) is the asymptotic interlayer conversion probability for a single channel with effective interlayer bias $\delta$ and tunneling $\lambda$, obtained by propagating the corresponding two-level Schrödinger equation from $t\to-\infty$ to $t\to+\infty$. For $p$-wave tunneling, $T_{\vec p} = T_1(p_x+ip_y)$, we define $\lambda_1 = T_1 \hbar k_{\mathrm{Z}}/E_{\mathrm{Z}}$, and the conductance $G$ can be calculated similarly as a function of $\delta$ and $\lambda_1$. 
\par

In Fig.~\ref{fig:conductance} we plot $G$, computed numerically, for both $s$-wave and $p$-wave tunneling. Two important features emerge: \textit{(i)} oscillations for $\delta<0$ as a function of $\delta$ and 
\textit{(ii)} a pronounced resonant ridge as a function of tunneling $\lambda$ or $\lambda_1$ where $G$ is maximized, the location of which depends weakly on $\delta$ (with the exception of $p$-wave for $\delta>0$). We contrast this with the simple, monotonic behavior $G \sim c_1\exp(- c_2 \delta^{3/2})$ expected for bulk semiconductors, for geometric factors $c_1,c_2$ and $\delta>0$~\cite{seabaugh2010low,ma2013interband}. 

\par 
The resonance \textit{(ii)} occurs at a characteristic field scale set by the interlayer tunneling. For $s$-wave tunneling at small $|\delta|$, it is $eF_{0}\approx 1.73 (2m^*/\hbar^2)^{1/2}\,T_0^{3/2}.$
For representative TMD parameters $m^*\sim 0.3$--$0.7\,m_e$ and $T_0\sim 1$--$20$~meV, this yields easily accessible fields $F_{0}\sim 10^5$--$10^7$~V/m. The interlayer bias $\Delta$ can be tuned independently via vertical gates, enabling systematic mapping of both signatures: sweeping $F$ through $F_{0}$ produces a conductance peak with width $O(F_{0})$, while varying $\Delta$ in the band overlap regime reveals the oscillations. We will show that these oscillations are periodic in $m^{\ast 1/2}|\Delta|^{3/2}/F$ at small fields. The physical origin of these features lies in quantum interference between distinct tunneling pathways, to which we turn next.

\paragraph*{Two-level scattering model.} Let us focus first on the case of $s$-wave tunneling. We consider $p_y=0$ without loss of generality, which can be absorbed by a shift of bandgap. Defining a dimensionless time $\tau = (p_x +eFt)/\hbar k_{\mathrm{Z}}$ and rescaling by $E_{\mathrm{Z}}$, the Hamiltonian reduces to the following two-level system: 
\begin{equation}\label{eq:Hoftau}
    H(\tau) = (\tau^2+\delta) \sigma^z + \lambda \sigma^x 
\end{equation}
where $\sigma^{z,x}$ are Pauli matrices. This is a quadratic generalization of the canonical Landau-Zener (LZ) problem $H_{\mathrm{LZ}}(\tau) = \tau\sigma^z + \lambda \sigma^x$, which has an exactly solvable asymptotic transition probability~\cite{landau1932theorie,zener1932non}. As in the LZ problem, a nonadiabatic transition is possible for Eq.~\eqref{eq:Hoftau}. \par
\begin{figure}
    \centering   \includegraphics[width=\linewidth]{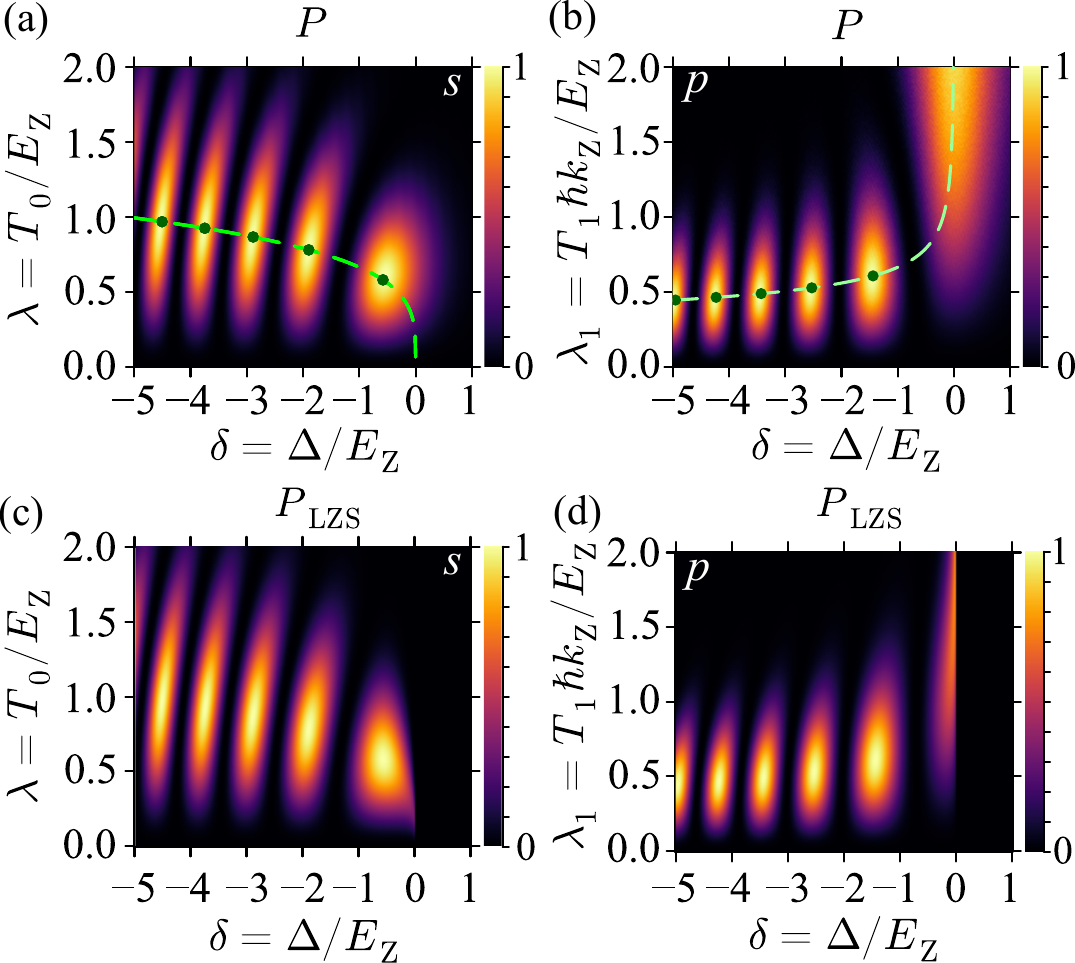}   
\caption{(a,b)~Tunneling probabilities for the two-level system Eq.~\eqref{eq:Hoftau} (and the $p$-wave analog), computed exactly. 
(c,d) Probabilities computed using the Landau-Zener-St\"uckelberg approximation, showing good agreement for $\Delta <0$. $P$ and $P_{\mathrm{LZS}}$ exhibit clear oscillations for $\Delta < 0$, a ridge of maximum probability, and isolated points of maximal transmission. The analytical predictions are superposed on (a,b): curves indicate  $P_{\text{LZ}} = 1/2$ (Eqs.~\eqref{eq:ridgeswave},\eqref{eq:ridgepwave}) and dots indicate perfect-transmission $P_{\text{LZS}} = 1$. For $p$-wave cases, $p_y=0$.}
\label{fig:probability}
\end{figure}

 
Let us define a propagator $U(\tau,\tau_0)$ by $U(\tau_0,\tau_0) = 1$ and
\begin{equation}
    i\hbar\partial_\tau U(\tau,\tau_0) = H(\tau) U(\tau,\tau_0).
\end{equation}
The transition probability occurring in Eq.~\eqref{eq:GmodeSum} is then
\begin{equation}\label{eq:Pdeltalambda}
    P(\delta,\lambda) = \lim_{\tau\to \infty}\lim_{\tau_0\to-\infty} \left| [U(\tau,\tau_0)]_{12}\right|^2.
\end{equation}
We plot $P(\delta,\lambda)$ in Fig.~\ref{fig:probability}a, computed numerically, where we observe that it captures the same qualitative features as the conductance $G$. The effect of $p_y$-integration in Eq.~\eqref{eq:GmodeSum} is simply a washing out of the sharp features.

The oscillations in $\delta$ are a manifestation of Landau-Zener-St\"uckelberg interferometry~\cite{shevchenko2010landau}, which describes the multiple-passage LZ problem. When $\delta <0$, the levels cross at two points $\pm \tau_0 = \pm\sqrt{-\delta}$, near which we may approximate Eq.~\eqref{eq:Hoftau} by $H(\tau)_{\pm}\approx \pm 2\tau_0 \tau \sigma^z + \lambda \sigma^x$. Between the two crossings, the evolution is approximately adiabatic, so that an eigenstate acquires a dynamical phase 
\begin{equation}
    \zeta(\delta,\lambda) = \int_{-\tau_0}^{\tau_0} \mathrm{d}\tau\, \sqrt{(\tau^2+\delta)^2+\lambda^2}.
\end{equation}
The final transition probability is given by~\cite{shevchenko2010landau}
\begin{equation}\label{eq:PfromLZS}
    P_{\mathrm{LZS}}(\delta,\lambda) = 4P_{\mathrm{LZ}}(1-P_{\mathrm{LZ}})\sin^2[\Phi_{\mathrm{St}}(\delta,\lambda)]
\end{equation}
where $P_{\mathrm{LZ}}(\delta,\lambda) = \exp(-2\pi\delta_{\mathrm{LZ}})$ with $\delta_{\mathrm{LZ}}
= \lambda^2/4\sqrt{-\delta}$ is the single-crossing LZ probability and
\begin{equation}
    \Phi_{\mathrm{St}}(\delta,\lambda)
    = \zeta(\delta,\lambda)    +\varphi_S(\delta_{\mathrm{LZ}})
\end{equation}
is the St\"uckelberg phase, defined in terms of the Stokes phase $\varphi_S(x) = \pi/4 +x(\ln x-1)+\arg \Gamma(1-ix)$. We show in Fig.~\ref{fig:probability}c that Eq.~\eqref{eq:PfromLZS} agrees reasonably well with $P$ with no fitting parameters, with agreement best for large, negative $\delta$. Outside that regime, the approximation of two well-separated crossings breaks down and $P_{\mathrm{LZS}}$ is not expected to approximate $P$. \par 

From the approximation Eq.~\eqref{eq:PfromLZS}, we can derive the following asymptotics. The period of oscillation in $\delta$ is $\Delta\delta\sim O(1)$ for $|\delta|\sim O(1)$, while $\Delta \delta \sim |\delta|^{-1/2}$ for large $|\delta|$ (so oscillations are a function of $\delta^{3/2}$). In physical terms, this corresponds to a period in $F$ scaling as $F^{2}/m^{\ast 1/2}|\Delta|^{3/2}$ for small fields, as shown in Fig.~\ref{fig:fig1}. Equivalently, the oscillations are periodic in $1/F$ at small fields, reminiscent of quantum oscillations in a magnetic field. The ridge of maximum probability follows the curve 
\begin{equation}\label{eq:ridgeswave}
    \lambda \sim \left(2\ln 2/\pi\right)^{1/2} (-\delta)^{1/4},
\end{equation}
obtained from setting $P_{\mathrm{LZ}}=1/2$. Along this ridge, when $\sin^2[\Phi_{\mathrm{St}}]=1$, there are isolated maxima at which $P_{\mathrm{LZS}}= 1$. For integers $n\gg 1$ these occur at 
\begin{equation}
    \left(\delta_n^*,\lambda_n^*\right) \sim \left(-(3\pi/4)^{2/3} n^{2/3},(6\ln^32/\pi^2)^{1/6}\, n^{1/6} \right).
\end{equation}
Remarkably, these are points of perfect transmission, made possible by perfect destructive interference of the amplitude for an electron to remain in its original band, analogous to a Fabry-Pérot interferometer, where unit transmission arises from destructive interference of the reflected wave. By contrast, in the traditional LZ problem, $P\to 1$ is only achieved at large tunneling. \par 

The case of $p$-wave tunneling can be worked out along similar lines. For $\Delta <0$, the dynamics again reduces to a double-passage Landau--Zener problem and exhibits
Landau--Zener--St\"uckelberg oscillations, but with an additional geometric phase and different scaling inherited from the linear-in-$p$ tunneling. Nevertheless, there is still a ridge of maximum probability satisfying 
\begin{equation}\label{eq:ridgepwave}
\lambda_1\approx (2\ln 2/\pi)^{1/2}(-\delta)^{-1/4},
\end{equation}
where we recall $\lambda_1= T_1\hbar k_{\rm Z}/{E_{\rm Z}}$, as well as points of perfect transmission, which for $n\gg 1$ occur at \begin{equation}
\left(\delta_n^\ast,\lambda_{1,n}^\ast\right) \sim \left( -(3\pi/4)^{2/3} n^{2/3} , (2^5\ln^32/3\pi^4)^{1/6}n^{-1/6}\right).
\end{equation}
We confirm the accuracy of these features in Fig.~\ref{fig:probability} and provide further details of derivations in~\cite{supp}. \par

Next, we turn to the non-monotonic behavior in the tunneling parameter $\lambda = T_0/E_{\mathrm{Z}}$. In particular, for any fixed gap parameter $\delta = \Delta/E_{\mathrm{Z}}$, including $\delta > 0$, the probability $P(\delta,\lambda)$ exhibits a global maximum at some $\lambda$, which we call the \textit{resonant peak}. This peak is exponentially suppressed in $\delta$ for $\delta >0$ but remains $O(1)$ otherwise. The existence of a resonant peak can be understood in simple energetic terms. When $\lambda \ll 1$, tunneling is weak and the transmission is near zero. On the other hand, when $\lambda \gg 1$, the two levels are split by $\sqrt{\delta^2+\lambda^2}\gg 1$, and transmission is once again highly suppressed. In between these limits, the transmission probability must have a maximum.\par 

A quantitative theory of the resonant peak can be derived for $\delta >0$ using the Dykhne-Davis-Pechukas (DDP) method~\cite{Dykhne1962,Davis1976Apr}, a saddle-point approximation summing over contributions from \textit{complex}-time avoided crossings. The DDP amplitude is an interfering sum over two complex-time saddle points $\tau^*$ corresponding to solutions of $\mathcal{E}(\tau^*)=0$ in the upper-half complex plane, where $\mathcal{E}(\tau) = 2\sqrt{(\tau^2+\delta)^2+\lambda^2}$ is the energy separation of $H(\tau)$. The probability is then
\begin{equation}\label{eq:PDDP}
    P_{\mathrm{DDP}}(\delta,\lambda)    \approx 4\,         e^{-2\,\Im\mathcal{D}(\delta,\lambda)}\,        \sin^2\!\bigl[\Re\mathcal{D}(\delta,\lambda)\bigr],
\end{equation}
where $\mathcal{D}(\delta,\lambda) = \int_0^{\tau^*} \mathcal{E}(\tau) \, d\tau$. The regime of validity is the adiabatic limit (large $\delta$), but the result agrees reasonably well with the exact result across the entire regime $\delta >0$, as shown in~\cite{supp}. For $\delta \to 0$, Eq.~\eqref{eq:PDDP} simplifies to
\begin{equation}  P_{\mathrm{DDP}}(\lambda) \sim 4e^{-2\sqrt{2}J\lambda^{3/2}}\sin^2(\sqrt{2}J\lambda^{3/2})
\end{equation}
where $J =\Gamma(1/4)\Gamma(3/2)/4\Gamma(7/4)$. The maximum is at $\lambda \approx 0.74$, reasonably close to the true maximum $\lambda\approx 0.69$ obtained numerically. We discuss the DDP approximation in further detail in~\cite{supp}. Thus across all regimes, the features of resonant Zener interferometry arise from the interference between (real- or complex-time) tunneling paths. \par 

\paragraph*{Experimental considerations.} The physical picture of two-level scattering sheds light on the phase-coherence time $t_\phi$ required for resonant Zener interferometry. For $\delta <0$, it is necessary for electrons to be coherent across the entire double-passage crossing. For $\delta_{\mathrm{LZ}}\gg 1$, each crossing takes time of order $(\lambda/|\delta|^{1/2})t_{\mathrm{Z}}$, while for $\delta_{\mathrm{LZ}}\ll 1$, each takes order $(1/|\delta|^{1/4})t_{\mathrm{Z}}$~\cite{PhysRevLett.62.2543}, while the time between crossings takes time of order $|\delta|^{1/2}t_{\mathrm{Z}}$ in both cases. For $\delta > 0$, the relevant time scale is $\max(\sqrt{\delta},\sqrt{\lambda}) t_{\mathrm{Z}}$: this is the time the electron spends in the region where tunneling is important compared to the level separation. Altogether, the phase-coherence time should satisfy
\begin{equation}\label{eq:tphi}
    \frac{t_\phi}{t_{\mathrm{Z}}} \gg \begin{cases}
       \mathcal{O}\left(|\delta|^{1/2} + \max[\lambda/|\delta|^{1/2},|\delta|^{-1/4}]\right)  & \delta < 0\\       \mathcal{O}\left(\max[\delta^{1/2},\lambda^{1/2}] \right)  & \delta >0
    \end{cases}.
\end{equation}
This criterion is always satisfied for large enough $F$. Moreover, for $F \sim 10^7$ V/m and $m^*\sim 0.3-0.7m_e$, one finds $t_{\mathrm{Z}}\sim 10-100$ fs, while typical relaxation and disorder-induced scattering times in hBN-encapsulated TMDs can exceed 1 ps at low temperatures~\cite{PhysRevB.99.115414,jiang2021interlayer}, comfortably satisfying Eq.~\eqref{eq:tphi}.
\par 
The analysis above assumes an approximately uniform in-plane field over a device region of length $L_x$. For the interferometric signatures to be observable, the field region must contain the full classically forbidden region where interlayer tunneling occurs, $L_x \gtrsim E_{\mathrm{gap}}/eF_{0}$. At the resonant fields $F_{0} \sim 10^5$--$10^7$~V/m and $E_{\mathrm{gap}}\sim 10$ meV, one needs $L_x \gtrsim 100$~nm, whereas devices can be up to micron-sized. Moreover, these fields remain well below the dielectric breakdown thresholds of TMDs and hBN ($\sim 10^8$--$10^9$~V/m), ensuring resonant Zener tunneling can be probed while maintaining device integrity. \par 

So far, we have not considered many-body effects. Interaction effects can be in part accounted for by band parameter renormalization. We may also consider the effect of attraction between the conduction electron and the valence hole left behind. One interesting possibility, which we set aside for now, is exciton formation. Another effect of the electron-hole interaction energy $U_{\mathrm{eh}}$ is a correction $\Delta \Phi$ to the dynamical phase in the Landau-Zener-St\"uckelberg interference. In \cite{supp} we calculate $\Delta \Phi$ and show that it is scales as $1/\sqrt{F}$ at low fields and as $1/F$ at high fields, and is hence negligible compared to the single-particle dynamical phase $\Phi_{\mathrm{St}}$ in both regimes. Hence, the salient features of resonant Zener interferometry are resilient to many-body effects. \par 

While our focus has been the effect of an in-plane electric field $F$, it is also interesting to consider an in-plane magnetic field $B$, which inserts flux between the layers. In \cite{supp} we show that $B$ effectively shifts $\Delta$ to
\begin{equation}
    \Delta(B) =\Delta + \frac{ e^2d^2B^2}{4\hbar^2 k_{\mathrm{Z}}^2}E_{\mathrm{Z}},
\end{equation}
where $d$ is the interlayer separation. Hence, for fixed $F>0$, in-plane magnetic field can also induce interferometric oscillations, with form given by replacing $\Delta$ with $\Delta(B)$ in our earlier analysis. With sufficient resolution, this may provide a new method to measure the interlayer separation $d$. Moreover, an in-plane $B$ can be used to shift the relative momentum offsets of conduction and valence bands in the event of imperfect band alignment. In summary, an in-plane magnetic field could prove a further useful experimental knob. 

\paragraph*{Discussion.} We have demonstrated that van der Waals heterostructures subject to in-plane electric fields realize a solid-state quantum interferometer exhibiting resonant Zener tunneling. Two key signatures emerge: Landau--Zener--St\"uckelberg oscillations in the band-overlap regime ($\Delta < 0$) periodic in $1/F$ at small fields and a pronounced resonance at $F_{0} \propto T_0^{3/2}$ in general. Both features occur robustly at experimentally accessible field scales and directly probe coherent interlayer charge-transfer dynamics.

These phenomena enable quantitative characterization of van der Waals heterostructures through purely electrical means. Sweeping the in-plane field through $F_{0}$ should yield a conductance peak whose location determines the interlayer tunneling strength $T_0$, while varying the interlayer bias $\Delta$ in the band-overlap regime produces oscillations whose periodicity relate to the effective mass $m^*$ and interlayer bias $\Delta$. These oscillations would be a noteworthy observation, as a new type of quantum oscillation depending on inverse \textit{electric} field rather than inverse magnetic field. Moreover, this effect could distinguish between $s$-wave and $p$-wave hybridization. In summary, resonant Zener interferometry could provide a useful characterization toolkit for device parameters which are typically extracted from optical spectroscopy or \textit{ab initio} calculations~\cite{wilson2017determination,hill2016band,PhysRevLett.121.026402}, complementing existing transport-based probes of two-dimensional materials. Moreover, it would manifest a nonrelativistic semiconductor-based analog of the Schwinger effect.

\par 
Beyond characterization, the nonmonotonic field dependence of the conductance opens pathways to new device applications. In particular, the current-voltage characteristic can exhibit negative differential resistance (NDR), which is exemplified by the Esaki tunnel diode~\cite{Nandkishore2011Aug,esaki1958new}. Here, it originates from interference effects between  forward-propagating Landau--Zener--St\"uckelberg tunneling, a qualitatively different mechanism. As a result, the NDR is primarily controlled by intrinsic parameters and is less sensitive to behavior of contacts. Moreover, the operation speed is set by the 10-100 fs Zener time instead of being limited by multiple backscatterings as in resonant-tunneling diodes~\cite{Nandkishore2011Aug,ng2007physics}.
\par 
Finally, the field scales for resonant Zener tunneling can lie below the ionization threshold for interlayer excitons in TMDs. The approximate condition $eF_0a_X \ll E_B$ to avoid field ionization, with $a_X$ and $E_B$ the Bohr radius and binding energy, gives $E_B/ea_X \sim 3 \times 10^7$ V/m for representative parameters ($E_B \sim 0.1$ eV, $a_X \sim 3$ nm), the far upper limit of the predicted typical $F_0$. This raises the prospect of field-tunable formation of long-lived interlayer excitons from Zener-tunneled electron-hole pairs, potentially unlocking a new avenue to controllable correlated excitonic phases~\cite{wilson2017determination,rivera2018interlayer,jiang2021interlayer,mak2022semiconductor,Zeng2023Nov,PhysRevB.92.165121}, a compelling direction for future work. 
\par 
\paragraph*{Acknowledgments.} We thank Jiaqi Cai for helpful discussions and Gal Shavit for valuable suggestions. NP acknowledges support
from the Walter Burke Institute for Theoretical Physics at Caltech. GR
is grateful for support from the Simons Foundation,
the Institute for Quantum Information and Matter,
an NSF Physics Frontiers Center (PHY-2317110), and
the AFOSR MURI program, under agreement number
FA9550-22-1-0339.

\bibliography{bib}

@article{qi2023thermodynamic,
  title={Thermodynamic behavior of correlated electron-hole fluids in van der Waals heterostructures},
  author={Qi, Ruishi and Joe, Andrew Y and Zhang, Zuocheng and Zeng, Yongxin and Zheng, Tiancheng and Feng, Qixin and Xie, Jingxu and Regan, Emma and Lu, Zheyu and Taniguchi, Takashi and others},
  journal={Nature communications},
  volume={14},
  number={1},
  pages={8264},
  year={2023},
  publisher={Nature Publishing Group UK London}
}

@article{PhysRevA.44.4280,
  title = {Interferences in adiabatic transition probabilities mediated by Stokes lines},
  author = {Joye, A. and Mileti, G. and Pfister, Ch.-Ed.},
  journal = {Phys. Rev. A},
  volume = {44},
  issue = {7},
  pages = {4280--4295},
  numpages = {0},
  year = {1991},
  month = {Oct},
  publisher = {American Physical Society},
  doi = {10.1103/PhysRevA.44.4280},
  url = {https://link.aps.org/doi/10.1103/PhysRevA.44.4280}
}

@article{shevchenko2010landau,
  title={Landau--zener--st{\"u}ckelberg interferometry},
  author={Shevchenko, Sergey N and Ashhab, Sahel and Nori, Franco},
  journal={Physics Reports},
  volume={492},
  number={1},
  pages={1--30},
  year={2010},
  publisher={Elsevier}
}

@article{Zeng2023Nov,
	author = {Zeng, Yongxin and Cr{\ifmmode\acute{e}\else\'{e}\fi}pel, Valentin and Millis, Andrew J.},
	title = {{Keldysh field theory of dynamical exciton condensation transitions in nonequilibrium electron-hole bilayers}},
	journal = {arXiv},
	year = {2023},
	month = nov,
	eprint = {2311.04074},
	doi = {10.1103/PhysRevLett.132.266001}
}

@article{PhysRevB.92.165121,
  title = {Theory of two-dimensional spatially indirect equilibrium exciton condensates},
  author = {Wu, Feng-Cheng and Xue, Fei and MacDonald, A. H.},
  journal = {Phys. Rev. B},
  volume = {92},
  issue = {16},
  pages = {165121},
  numpages = {13},
  year = {2015},
  month = {Oct},
  publisher = {American Physical Society},
  doi = {10.1103/PhysRevB.92.165121},
  url = {https://link.aps.org/doi/10.1103/PhysRevB.92.165121}
}

@article{Vasilev2014Feb,
	author = {Vasilev, G. S. and Vitanov, N. V.},
	title = {{Coherent excitation of two-state system by a Lorentzian filed}},
	journal = {arXiv},
	year = {2014},
	month = feb,
	eprint = {1402.5119},
	doi = {10.48550/arXiv.1402.5119}
}

@article{Schwinger1951Jun,
	author = {Schwinger, Julian},
	title = {{On Gauge Invariance and Vacuum Polarization}},
	journal = {Phys. Rev.},
	volume = {82},
	number = {5},
	pages = {664--679},
	year = {1951},
	month = jun,
	publisher = {American Physical Society},
	doi = {10.1103/PhysRev.82.664}
}

@article{Davis1976Apr,
	author = {Davis, Jon P. and Pechukas, Philip},
	title = {{Nonadiabatic transitions induced by a time{-}dependent Hamiltonian in the semiclassical/adiabatic limit: The two{-}state case}},
	journal = {J. Chem. Phys.},
	volume = {64},
	number = {8},
	pages = {3129--3137},
	year = {1976},
	month = apr,
	issn = {0021-9606},
	publisher = {AIP Publishing},
	doi = {10.1063/1.432648}
}

@article{Dykhne1962,
	author = {Dykhne, Aleksandr M Dykhne},
	title = {{Adiabatic Perturbation of Discrete Spectrum States }},
	journal = {Soviet Physics JETP },
	volume = {14},
	number = {4},
	year = {1962},
	month = apr,
}

@article{Rivera2018Nov,
	author = {Rivera, Pasqual and Yu, Hongyi and Seyler, Kyle L. and Wilson, Nathan P. and Yao, Wang and Xu, Xiaodong},
	title = {{Interlayer valley excitons in heterobilayers of transition metal dichalcogenides}},
	journal = {Nat. Nanotechnol.},
	volume = {13},
	pages = {1004--1015},
	year = {2018},
	month = nov,
	issn = {1748-3395},
	publisher = {Nature Publishing Group},
	doi = {10.1038/s41565-018-0193-0}
}

@misc{supp,
  note = "See the Supplemental Material at for further details."
}

@article{Nandkishore2011Aug,
	author = {Nandkishore, Rahul and Levitov, Leonid},
	title = {{Common-path interference and oscillatory Zener tunneling in bilayer graphene p-n junctions}},
	journal = {Proc. Natl. Acad. Sci. U.S.A.},
	volume = {108},
	number = {34},
	pages = {14021--14025},
	year = {2011},
	month = aug,
	publisher = {Proceedings of the National Academy of Sciences},
	doi = {10.1073/pnas.1101352108}
}

@article{mak2022semiconductor,
  title={Semiconductor moir{\'e} materials},
  author={Mak, Kin Fai and Shan, Jie},
  journal={Nature Nanotechnology},
  volume={17},
  number={7},
  pages={686--695},
  year={2022},
  publisher={Nature Publishing Group UK London}
}

@article{landau1932theorie,
  title={Zur theorie der energieubertragung. II},
  author={Landau, Lev},
  journal={Physikalische Zeitschrift der Sowjetunion},
  volume={2},
  pages={46},
  year={1932}
}

@article{scammell2025zener,
  title={Zener tunnelling in biased bilayer graphene via analytic continuation of semiclassical theory},
  author={Scammell, Harley and Sushkov, Oleg P},
  journal={arXiv preprint arXiv:2505.24150},
  year={2025}
}

@article{jiang2021interlayer,
  title={Interlayer exciton formation, relaxation, and transport in TMD van der Waals heterostructures},
  author={Jiang, Ying and Chen, Shula and Zheng, Weihao and Zheng, Biyuan and Pan, Anlian},
  journal={Light: Science \& Applications},
  volume={10},
  number={1},
  pages={72},
  year={2021},
  publisher={Nature Publishing Group UK London}
}

@book{ng2007physics,
  title={Physics of semiconductor devices},
  author={Ng, Kwok Kwok and Sze, Simon M},
  year={2007},
  publisher={Wiley-Interscience Hoboken, NJ}
}

@article{esaki1958new,
  title={New phenomenon in narrow germanium p- n junctions},
  author={Esaki, Leo},
  journal={Physical review},
  volume={109},
  number={2},
  pages={603},
  year={1958},
  publisher={APS}
}

@article{PhysRevB.99.115414,
  title = {Weak localization in boron nitride encapsulated bilayer ${\mathrm{MoS}}_{2}$},
  author = {Papadopoulos, Nikos and Watanabe, Kenji and Taniguchi, Takashi and van der Zant, Herre S. J. and Steele, Gary A.},
  journal = {Phys. Rev. B},
  volume = {99},
  issue = {11},
  pages = {115414},
  numpages = {5},
  year = {2019},
  month = {Mar},
  publisher = {American Physical Society},
  doi = {10.1103/PhysRevB.99.115414},
  url = {https://link.aps.org/doi/10.1103/PhysRevB.99.115414}
}

@article{PhysRevLett.62.2543,
  title = {Time of Zener tunneling},
  author = {Mullen, Kieran and Ben-Jacob, Eshel and Gefen, Yuval and Schuss, Zeev},
  journal = {Phys. Rev. Lett.},
  volume = {62},
  issue = {21},
  pages = {2543--2546},
  numpages = {0},
  year = {1989},
  month = {May},
  publisher = {American Physical Society},
  doi = {10.1103/PhysRevLett.62.2543},
  url = {https://link.aps.org/doi/10.1103/PhysRevLett.62.2543}
}

@article{landauer1957spatial,
  title={Spatial variation of currents and fields due to localized scatterers in metallic conduction},
  author={Landauer, Rolf},
  journal={IBM Journal of research and development},
  volume={1},
  number={3},
  pages={223--231},
  year={1957},
  publisher={IBM}
}

@article{schmitt2023mesoscopic,
  title={Mesoscopic Klein-Schwinger effect in graphene},
  author={Schmitt, Aur{\'e}lien and Vallet, Pierre and Mele, David and Rosticher, Micha{\"e}l and Taniguchi, Takashi and Watanabe, Kenji and Bocquillon, Erwan and Feve, Gwendal and Berroir, Jean-Marc and Voisin, Christophe and others},
  journal={Nature Physics},
  volume={19},
  number={6},
  pages={830--835},
  year={2023},
  publisher={Nature Publishing Group UK London}
}

@article{sauter1931verhalten,
  title={{\"U}ber das Verhalten eines Elektrons im homogenen elektrischen Feld nach der relativistischen Theorie Diracs},
  author={Sauter, Fritz},
  journal={Zeitschrift f{\"u}r Physik},
  volume={69},
  number={11},
  pages={742--764},
  year={1931},
  publisher={Springer}
}

@article{daus2021high,
  title={High-performance flexible nanoscale transistors based on transition metal dichalcogenides},
  author={Daus, Alwin and Vaziri, Sam and Chen, Victoria and Koroglu, Cagil and Grady, Ryan W and Bailey, Connor S and Lee, Hye Ryoung and Schauble, Kirstin and Brenner, Kevin and Pop, Eric},
  journal={Nature Electronics},
  volume={4},
  number={7},
  pages={495--501},
  year={2021},
  publisher={Nature Publishing Group UK London}
}

@article{hill2016band,
  title={Band alignment in MoS2/WS2 transition metal dichalcogenide heterostructures probed by scanning tunneling microscopy and spectroscopy},
  author={Hill, Heather M and Rigosi, Albert F and Rim, Kwang Taeg and Flynn, George W and Heinz, Tony F},
  journal={Nano letters},
  volume={16},
  number={8},
  pages={4831--4837},
  year={2016},
  publisher={ACS Publications}
}

@article{wilson2017determination,
  title={Determination of band offsets, hybridization, and exciton binding in 2D semiconductor heterostructures},
  author={Wilson, Neil R and Nguyen, Paul V and Seyler, Kyle and Rivera, Pasqual and Marsden, Alexander J and Laker, Zachary PL and Constantinescu, Gabriel C and Kandyba, Viktor and Barinov, Alexei and Hine, Nicholas DM and others},
  journal={Science advances},
  volume={3},
  number={2},
  pages={e1601832},
  year={2017},
  publisher={American Association for the Advancement of Science}
}

@article{zener1932non,
  title={Non-adiabatic crossing of energy levels},
  author={Zener, Clarence},
  journal={Proceedings of the Royal Society of London. Series A, Containing Papers of a Mathematical and Physical Character},
  volume={137},
  number={833},
  pages={696--702},
  year={1932},
  publisher={The Royal Society London}
}

@article{devakul2021magic,
  title={Magic in twisted transition metal dichalcogenide bilayers},
  author={Devakul, Trithep and Cr{\'e}pel, Valentin and Zhang, Yang and Fu, Liang},
  journal={Nature communications},
  volume={12},
  number={1},
  pages={6730},
  year={2021},
  publisher={Nature Publishing Group UK London}
}

@article{PhysRevLett.121.026402,
  title = {Hubbard Model Physics in Transition Metal Dichalcogenide Moir\'e Bands},
  author = {Wu, Fengcheng and Lovorn, Timothy and Tutuc, Emanuel and MacDonald, A. H.},
  journal = {Phys. Rev. Lett.},
  volume = {121},
  issue = {2},
  pages = {026402},
  numpages = {5},
  year = {2018},
  month = {Jul},
  publisher = {American Physical Society},
  doi = {10.1103/PhysRevLett.121.026402},
  url = {https://link.aps.org/doi/10.1103/PhysRevLett.121.026402}
}

@article{rivera2018interlayer,
  title={Interlayer valley excitons in heterobilayers of transition metal dichalcogenides},
  author={Rivera, Pasqual and Yu, Hongyi and Seyler, Kyle L and Wilson, Nathan P and Yao, Wang and Xu, Xiaodong},
  journal={Nature nanotechnology},
  volume={13},
  number={11},
  pages={1004--1015},
  year={2018},
  publisher={Nature Publishing Group UK London}
}

@article{regan2022emerging,
  title={Emerging exciton physics in transition metal dichalcogenide heterobilayers},
  author={Regan, Emma C and Wang, Danqing and Paik, Eunice Y and Zeng, Yongxin and Zhang, Long and Zhu, Jihang and MacDonald, Allan H and Deng, Hui and Wang, Feng},
  journal={Nature Reviews Materials},
  volume={7},
  number={10},
  pages={778--795},
  year={2022},
  publisher={Nature Publishing Group UK London}
}

@article{castellanos2022van,
  title={Van der Waals heterostructures},
  author={Castellanos-Gomez, Andres and Duan, Xiangfeng and Fei, Zhe and Gutierrez, Humberto Rodriguez and Huang, Yuan and Huang, Xinyu and Quereda, Jorge and Qian, Qi and Sutter, Eli and Sutter, Peter},
  journal={Nature Reviews Methods Primers},
  volume={2},
  number={1},
  pages={58},
  year={2022},
  publisher={Nature Publishing Group UK London}
}

@article{novoselov20162d,
  title={2D materials and van der Waals heterostructures},
  author={Novoselov, K Sꎬ and Mishchenko, Artem and Carvalho, Alexandra and Castro Neto, AH},
  journal={Science},
  volume={353},
  number={6298},
  pages={aac9439},
  year={2016},
  publisher={American Association for the Advancement of Science}
}

@article{geim2013van,
  title={Van der Waals heterostructures},
  author={Geim, Andre K and Grigorieva, Irina V},
  journal={Nature},
  volume={499},
  number={7459},
  pages={419--425},
  year={2013},
  publisher={Nature Publishing Group UK London}
}

@article{PhysRevB.74.041403,
  title = {Selective transmission of Dirac electrons and ballistic magnetoresistance of $n\text{\ensuremath{-}}p$ junctions in graphene},
  author = {Cheianov, Vadim V. and Fal'ko, Vladimir I.},
  journal = {Phys. Rev. B},
  volume = {74},
  issue = {4},
  pages = {041403},
  numpages = {4},
  year = {2006},
  month = {Jul},
  publisher = {American Physical Society},
  doi = {10.1103/PhysRevB.74.041403},
  url = {https://link.aps.org/doi/10.1103/PhysRevB.74.041403}
}

@article{johri2013common,
  title = {Common path interference in Zener tunneling is a universal phenomenon},
  author = {Johri, Sonika and Nandkishore, Rahul and Bhatt, R. N. and Mele, E. J.},
  journal = {Phys. Rev. B},
  volume = {87},
  issue = {23},
  pages = {235413},
  numpages = {7},
  year = {2013},
  month = {Jun},
  publisher = {American Physical Society},
  doi = {10.1103/PhysRevB.87.235413},
  url = {https://link.aps.org/doi/10.1103/PhysRevB.87.235413}
}

@inproceedings{ma2013interband,
  title={Interband tunneling transport in 2-dimensional crystal semiconductors},
  author={Ma, Nan and Jena, Debdeep},
  booktitle={71st Device Research Conference},
  pages={103--104},
  year={2013},
  organization={IEEE}
}

@article{seabaugh2010low,
  title={Low-voltage tunnel transistors for beyond CMOS logic},
  author={Seabaugh, Alan C and Zhang, Qin},
  journal={Proceedings of the IEEE},
  volume={98},
  number={12},
  pages={2095--2110},
  year={2010},
  publisher={IEEE}
}

@article{zener1934theory,
  title={A theory of the electrical breakdown of solid dielectrics},
  author={Zener, Clarence},
  journal={Proceedings of the Royal Society of London. Series A, Containing Papers of a Mathematical and Physical Character},
  volume={145},
  number={855},
  pages={523--529},
  year={1934},
  publisher={The Royal Society London}
}

\end{document}


\title{Supplemental Material: Resonant Zener Interferometry in van der Waals Heterostructures}

\author{Nisarga Paul}
\email{npaul@caltech.edu}
\affiliation{Department of Physics and Institute for Quantum Information and Matter,
California Institute of Technology, Pasadena, California 91125, USA}
\author{Gil Refael}
\email{refael@caltech.edu}
\affiliation{Department of Physics and Institute for Quantum Information and Matter,
California Institute of Technology, Pasadena, California 91125, USA}

\maketitle

\tableofcontents

\section{Details of $s$-wave model}

We present details of the approximations presented for minimal model with $s$-wave tunneling. First, we recall the definitions: $H(t)=H_0(\vec p+e\vec A(t))$ with $\vec A(t)=-Ft\,\hat x$ and
\begin{equation}
    H_0(\vec p)=
    \begin{pmatrix}
        p^2/2m^*+\Delta & T_0\\
        T_0 & -p^2/2m^*-\Delta
    \end{pmatrix},
\end{equation}
where $\vec p=(p_x,p_y)$, $m^*$ is effective mass, $\Delta$ is the interlayer bias, and
$T_0$ is the interlayer tunneling. Define the Zener length and energy scales
\begin{equation}
    \ell_{\mathrm Z} =\left(\hbar^2/2m^*eF\right)^{1/3}= k_\mathrm{Z}^{-1},\quad
    E_{\mathrm Z} =eF\ell_{\mathrm{Z}}= \hbar t_{\mathrm{Z}}^{-1}.
\end{equation}
Focusing on $p_y=0$ (by shifting $\Delta$), defining dimensionless time $\tau = (p_x +eFt)/\hbar k_{\mathrm{Z}}$ and rescaling by $E_{\mathrm{Z}}$, the Hamiltonian reduces to 
\begin{equation}
    H(\tau) = (\tau^2+\delta) \sigma^z + \lambda \sigma^x 
\end{equation}
where $\sigma^{z,x}$ are Pauli matrices and $\delta=\Delta/E_{\mathrm Z},
\lambda=T_0/E_{\mathrm Z}$. We take $U(\tau,\tau_0)$ to solve $i\hbar\partial_\tau U(\tau,\tau_0) = H(\tau) U(\tau,\tau_0)$
with $U(\tau_0,\tau_0) = 1$, and define the transition probability
\begin{equation}
    P(\delta,\lambda) = \lim_{\tau\to \infty}\lim_{\tau_0\to-\infty} \left| [U(\tau,\tau_0)]_{12}\right|^2.
\end{equation}

\begin{figure}
    \centering   \includegraphics[width=0.65\linewidth]{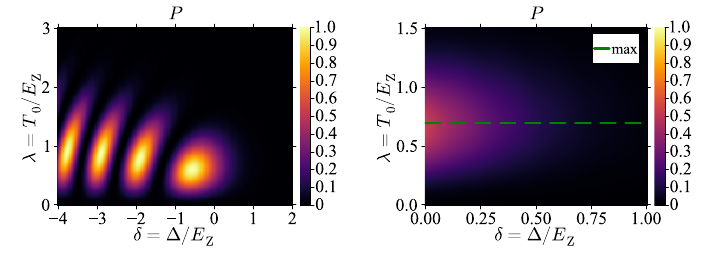}   
    \caption{ \textbf{Probability for $s$-wave tunneling.} (Left) For a large range of $\delta,\lambda$. (Right) For a limited range $0<\delta<1,0<\lambda<1.5$. In this regime, maximal $P$ is achieved at $\lambda \approx 0.6935$, shown in dashed line.}
    \label{fig:tunnelingheatmap}
\end{figure}

\subsection{Landau-Zener-St\"uckelberg approximation}
\begin{figure}[t!]
    \centering   \includegraphics[width=0.65\linewidth]{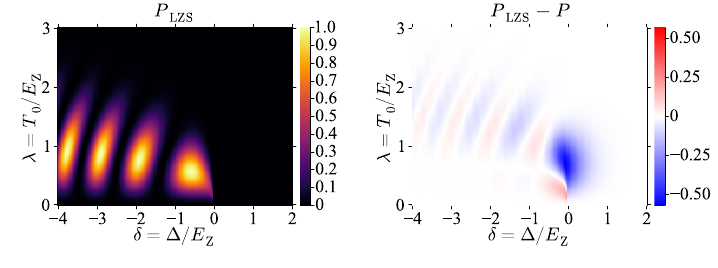}   
    \caption{\textbf{Landau-Zener-St\"uckelberg (LZS) probability.} (Left) $P_{\mathrm{LZS}}$, computed from the LZS approximation \eqref{eq:PLZStuckelberg}. (Right) Difference with exact $P$. Good agreement is found for lower $\delta$, the regime of validity. }
 \label{fig:tunnelingheatmapLZ}
\end{figure}

When $\delta < 0$, $H(\tau)$ describes a double–passage Landau–Zener (LZ) problem~\cite{shevchenko2010landau}. We linearize the Hamiltonian in the vicinity of each
crossing $\pm \tau_0 = \pm\sqrt{-\delta}$. Near either crossing
\begin{equation}
    H(\tau)_{\pm} \simeq \pm 2\tau_0\,\tau\,\sigma^z + \lambda\,\sigma^x.
\end{equation}
This is the standard Landau-Zener problem with nonadiabatic single–pass transition probability
\begin{equation}
    P_{\mathrm{LZ}}(\delta,\lambda)
    = \exp\!\bigl[-2\pi\delta_{\mathrm{LZ}}\bigr],\qquad \delta_{\mathrm{LZ}}
    = \frac{\lambda^2}{4\sqrt{-\delta}}.
\end{equation}
Between the two crossings, the evolution is approximately
adiabatic, so that an eigenstate acquires a dynamical phase
\begin{equation}
    \zeta(\delta,\lambda)
    = \int_{-\tau_0}^{\tau_0}
d\tau\;\sqrt{\bigl(\tau^2+\delta\bigr)^2 + \lambda^2}.
\end{equation}
Each LZ passage is described in the adiabatic basis by the
unitary ``impulse'' matrix~\cite{shevchenko2010landau}
\begin{equation}
    N =
    \begin{pmatrix}
        \sqrt{1-P_{\mathrm{LZ}}}\,e^{-i\tilde\varphi_S} & -\sqrt{P_{\mathrm{LZ}}} \\
        \sqrt{P_{\mathrm{LZ}}} & \sqrt{1-P_{\mathrm{LZ}}}\,e^{+i\tilde\varphi_S}
    \end{pmatrix},
\end{equation}
where the Stokes phase is $\tilde\varphi_S = \varphi_S -\frac{\pi}{2}$ with
\begin{equation}
    \varphi_S = \frac{\pi}{4} +\delta_{\mathrm{LZ}}(\ln \delta_{\mathrm{LZ}}-1)+\arg \Gamma(1-i\delta_{\mathrm{LZ}}).
\end{equation}
 Starting at $\tau\to -\infty$ in the lower adiabatic band, the
sequence LZ passage $\to$ adiabatic evolution $\to$ LZ passage leads to
the double–pass transition amplitude
\begin{equation}
    \mathcal{A}_{1\to 2}(\delta,\lambda)
    = -\,2\sqrt{P_{\mathrm{LZ}}(1-P_{\mathrm{LZ}})}\,
      \sin\!\bigl[\Phi_{\mathrm{St}}(\delta,\lambda)\bigr],
\end{equation}
where the Stückelberg phase, defined
\begin{equation}
    \Phi_{\mathrm{St}}(\delta,\lambda)
    = \zeta(\delta,\lambda)
      +\varphi_S\!\bigl(\delta_{\mathrm{LZ}}(\delta,\lambda)\bigr),
\end{equation}
captures both dynamical and Stokes contributions.  The
corresponding double–pass transition probability is
\begin{equation}\label{eq:PLZStuckelberg}
    P_{\rm LZS}(\delta,\lambda)
    \simeq 4 P_{\mathrm{LZ}}(1-P_{\mathrm{LZ}})\,      \sin^2\!\bigl[\Phi_{\mathrm{St}}(\delta,\lambda)\bigr].
\end{equation}
We plot this in Fig.~\ref{fig:tunnelingheatmapLZ}, along with its difference with the exact numerical result. 
Good agreement is found in general with no fitting parameters.
Agreement is best for lower $\delta$ where the LZ transitions are well-separated, and poorer agreement is found near $\delta =0$, which is expected due to breakdown of the assumption of independent LZ transitions. \par

The period of oscillations is $O(1)$ when $|\delta|$ is $O(1)$. The period also takes a simple form for large $|\delta|$. In this regime, $\varphi_S \to \pi/4$ and $\zeta(\delta,\lambda) \approx\frac43 |\delta|^{3/2}$, so 
$
    \Phi_{\mathrm{St}} \approx \frac{\pi}{4} + \frac43|\delta|^{3/2}.
$
Then the period $\Delta \delta$ scales as
\begin{equation}
    \Delta\delta \approx \frac{\pi}{2\sqrt{|\delta|}}.
\end{equation}
Notably, the two-level system exhibits isolated resonances of \textit{perfect} transmission where the incoming valence electron scatters into the conduction band with unit probability, reminiscent of the transmission peaks of a Fabry-P\'erot interferometer but arising solely due to forward propagating waves. By way of contrast, this is only achieved for the single-crossing LZ problem asymptotically. From Eq.~\eqref{eq:PLZStuckelberg}, it is evident that this is achieved only when two conditions are met simultaneously: the partial Landau-Zener transitions must satisfy $P_{\mathrm{LZ}}=1/2$, and the accumulated phase must be constructive ($\Phi_{\mathrm{St}} = (n+\frac{1}{2})\pi$).

The amplitude condition $P_{\mathrm{LZ}}=1/2$ implies the relation
$
    \lambda^4 = \left(\frac{2\ln 2}{\pi}\right)^2 |\delta|.
$
The phase condition implies $|\delta| \sim \frac{1}{4} (3\pi/2)^{2/3}(4n+1)^{2/3}$ at large $|\delta|$. Together these identify the resonances of perfect transmission as the result quoted in the main text, approximately $(\delta_n^*, \lambda_n^*) \sim (-1.77\,n^{2/3},0.77\,n^{1/6})$ for large integers $n\gg 1$. These results were confirmed in Fig. 3 of the main text. 

\subsection{DDP method, $\delta>0$}

When $\delta > 0$, \textit{i.e.} there is a bandgap, a different analytical approach is necessary. Numerically, we find that the maximum amplitude is achieved at approximately constant value of $\lambda\approx 0.6935$, corresponding to the condition 
\begin{equation}
    T_0^3\approx 0.33\frac{\hbar^2}{2m^*}F^2,
\end{equation}
as shown in Fig.~\ref{fig:tunnelingheatmap}. In this regime conduction and valence bands are separated by a gap and there is no real crossing of the adiabatic levels. Hence the Landau--Zener picture is not directly applicable, and we instead employ the Dykhne--Davis--Pechukas (DDP) method~\cite{Dykhne1962,Davis1976Apr}, a saddle-point method summing over contributions from complex-time avoided crossings.
\par 
Without loss of generality we again set $\kappa_y=0$ and work with the dimensionless two-level Hamiltonian $H(\tau) = \bigl(\tau^2+\delta\bigr)\sigma^z + \lambda\,\sigma^x$ with instantaneous energy level splitting
\begin{equation}
    \mathcal{E}(\tau) = 2\sqrt{\bigl(\tau^2+\delta\bigr)^2 + \lambda^2}.
\end{equation}
We must find solutions to $\mathcal{E}(\tau^*)=0$ in the upper-half of the complex plane. There are two, namely
\begin{equation} \tau^*_{\pm} = \pm \sqrt{-\delta \pm i\lambda}
\end{equation}
The transition probability in the DDP approximation is 
\begin{equation} P_{\rm DDP}(\delta,\lambda)  \approx \left| \sum_{s=\pm} \Gamma_s e^{i\mathcal{D}(\tau^*_s)}\right|^2
\end{equation}
with $\Gamma_s = 4i\lim_{\tau\to \tau_s^*}(\tau-\tau_s^*)\dot\vartheta(\tau) = \mp 1$ where $\tan 2\vartheta(\tau) = -\lambda/ (\tau^2+\delta)$~\cite{Davis1976Apr,Vasilev2014Feb} and
\begin{equation}
    \mathcal{D}(\tau^*)
    = \int_0^{\tau^*} \mathcal{E}(\tau)\,d\tau
\end{equation}
is a complex action evaluated along a contour from $0$ to $\tau^*$. Writing $\mathcal{D}(\delta,\lambda)
    \equiv \mathcal{D}(\tau^*_+)$,
the probability can be written as
\begin{equation}\label{eq:DDPProb}
   P_{\rm DDP}(\delta,\lambda) 
    \approx 4\,
    e^{-2\,\Im\mathcal{D}(\delta,\lambda)}\,        \sin^2\!\bigl[\Re\mathcal{D}(\delta,\lambda)\bigr],
\end{equation}
Though this is generally difficult to simplify further, for $\delta=0$ we can proceed analytically. We have
$\tau_+^* = \lambda^{1/2} e^{i\pi/4}$
and
\begin{equation}
    \mathcal{D}(0,\lambda)
    = 2\int_0^{\tau_+^*} d\tau\,\sqrt{\tau^4+\lambda^2}
    = 2\lambda^{3/2} e^{i\pi/4} J,
\end{equation}
with $J = \int_0^1\!dt\,\sqrt{1-t^4}
      = \frac{1}{4}\,\frac{\Gamma(1/4)\Gamma(3/2)}{\Gamma(7/4)}.$
Then Eq.~\eqref{eq:DDPProb} gives
\begin{equation}
    P_{\rm DDP}(0,\lambda) 
    \approx 4\exp\!\bigl[-2\sqrt{2}J\,\lambda^{3/2}\bigr]\,
           \sin^2\!\bigl(\sqrt{2}J\,\lambda^{3/2}\bigr).
\end{equation}
which has a maximum at $\lambda \approx 0.74$.\par
\begin{figure}
    \centering
    \includegraphics[width=0.65\linewidth]{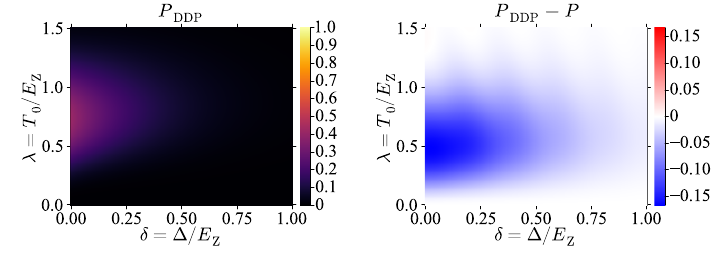}

    \caption{\textbf{Dykhne-Davis-Pechukas (DDP) probability.} (Left) DDP probability with no fitting parameters (Eq. \eqref{eq:DDPProb}). Maximal $P_{\mathrm{DDP}}$ occurs at $\lambda \approx 0.79$. (Right) difference with exact $P$.}
    \label{fig:DDP}
\end{figure}

 
In general we must evaluate $\mathcal{D}(\delta,\lambda)$ numerically and resulting DDP approximation~\eqref{eq:DDPProb} reproduces the main features of the exact numerical $P(\delta,\lambda)$ in the $\delta>0$ regime. Quantitatively, the DDP approximation agrees with the full time-dependent solution at the level of $|\Delta P|\lesssim 0.15$ in the region where $P(\delta,\lambda)$ is appreciable, and predicts the location of the maximum at $\lambda\approx 0.8$, reasonably close to the exact value $\lambda\approx 0.7$ extracted from Fig.~\ref{fig:tunnelingheatmap} by fitting over $\delta>0$. We plot the DDP result in Fig.~\ref{fig:DDP}. \par 

The DDP approximation is asymptotically exact in the adiabatic limit for Hamiltonians whose energies satisfying some technical conditions on their complex zeros~\cite{PhysRevA.44.4280} satisfied here. The adiabatic limit corresponds to weak fields, hence large $\delta,\lambda$. This is consistent with Fig.~\ref{fig:DDP}. Remarkably, however, the DDP approximation is still very good to small $\delta,\lambda$ and reproduces the correct qualitative behavior there as well, consistent with observations of \cite{PhysRevA.44.4280,Vasilev2014Feb}.

\section{Details of $p$-wave model}
\label{sec:pwave}

We consider the case of pure $p$-wave tunneling, 
\begin{equation}
    T_{\vec p} = T_1 (p_x + i p_y).
\end{equation}
We omit the valley index because integration over momenta results in a valley-independent conductance. Once again we use the Zener field scales
\begin{equation}   \ell_\mathrm{Z}=\left(\frac{\hbar^2}{2m^*F}\right)^{1/3} = k_{\mathrm{Z}}^{-1},\quad 
E_Z=\frac{\hbar^2}{2m^*\ell_Z^2}=F\ell_Z = \hbar \tau_{\mathrm{Z}}^{-1}
\end{equation}
and define dimensionless variables
\begin{equation}
    \tau= \frac{t}{\tau_\mathrm{Z}},    \quad\vec \kappa= \frac{\vec k}{\hbar k_\mathrm{Z}},
\quad
    \delta= \frac{\Delta}{E_\mathrm{Z}},\quad
    \lambda_1= \frac{T_1\hbar k_\mathrm{Z}}{E_\mathrm{Z}}.
\end{equation}
In these units, rescaled by $E_\mathrm{Z}$, the Hamiltonian reads
\begin{equation}
H(\tau)    =\Big[(\kappa_x+\tau)^2+\kappa_y^2+\delta\Big]\sigma^z     +\lambda_1\Big[(\kappa_x+\tau)\sigma^x-\kappa_y\sigma^y\Big],
    \label{eq:h_pwave_dimless}
\end{equation}
and the asymptotic band-conversion probability is obtained from the evolution from $\tau=-\infty$ to
$\tau=+\infty$,
\begin{equation}
    P_{\vec\kappa}(\delta,\lambda)
    = \big|\,[U_{\vec\kappa}(+\infty,-\infty)]_{cv}\big|^2,
\end{equation}
where $U_{\vec \kappa}$ solves $i\hbar \partial_{\tau}U_{\vec\kappa}(\tau,\tau_0)=H(\tau)U_{\vec \kappa}(\tau,\tau_0)$.
As in the $s$-wave case, $\kappa_x$ merely fixes the time origin; shifting $\tau\to\tau-\kappa_x$
allows one to set $\kappa_x=0$ without loss of generality. In contrast to $s$-wave tunneling, the
transverse momentum cannot be absorbed into $\delta$, because $\kappa_y$ enters both the detuning and
the transverse coupling $\kappa_y\sigma^y$.

\subsection{Landau--Zener--St\"uckelberg, ($\delta+\kappa_y^2<0$)}
\label{subsec:pwave_LZS}

Setting $\kappa_x=0$, the dimensionless Hamiltonian becomes
\begin{equation}
    H(\tau)
    =(\tau^2+\delta+\kappa_y^2)\sigma^z
     +\lambda_1\big(\tau\sigma^x-\kappa_y\sigma^y\big).
    \label{eq:h_pwave_kx0}
\end{equation}
When $\delta+\kappa_y^2<0$ the energy levels undergo two avoided crossings, at
\begin{equation}
    \tau=\pm\tau_0,\quad \tau_0=\sqrt{-\delta-\kappa_y^2}.
\end{equation}
Linearizing about either crossing gives a standard
Landau--Zener problem with slope $2\tau_0$ and a complex transverse coupling with magnitude
$\Delta_\perp=\lambda_1\sqrt{\tau_0^2+\kappa_y^2}=\lambda_1\sqrt{-\delta}.$
The single-pass Landau--Zener probability is
\begin{equation}
    P_{\rm LZ}(\delta,\lambda_1)
    =\exp(-2\pi \delta_{\mathrm{LZ}}),\quad \delta_{\mathrm{LZ}} = \frac{ \Delta_\perp^2}{4\tau_0}.
    \label{eq:PLZ_pwave}
\end{equation}
Between the two crossings the evolution is approximately adiabatic and the state acquires the
dynamical phase
\begin{equation}
    \zeta(\delta,\lambda_1)
    =\int_{-\tau_0}^{\tau_0} d\tau\;
      \sqrt{(\tau^2-\tau_0^2)^2+\lambda_1^2(\tau^2+\kappa_y^2)}.
    \label{eq:zeta_pwave}
\end{equation}
Combining the two LZ passages in the standard adiabatic--impulse picture yields the usual
Landau--Zener--St\"uckelberg form
\begin{equation}
    P(\delta,\lambda_1)
    \simeq 4P_{\rm LZ}(1-P_{\rm LZ})   \sin^2\!\Big[\zeta(\delta,\lambda_1)+\varphi_S(\delta_{\rm LZ})+\phi_g\Big],  \label{eq:LZS_pwave}
\end{equation}
where $\varphi_S$ is the standard Stokes phase~\cite{shevchenko2010landau} and
\begin{equation}
\phi_g
    = \frac12\,\arg\!\left(\frac{-\tau_0+i\kappa_y}{\tau_0+i\kappa_y}\right)
\end{equation}
is an additional phase originating from the chiral (complex) $p$-wave tunneling matrix element.
\par 
Within the LZS approximation \eqref{eq:LZS_pwave}, local maxima of the channel-resolved conversion
probability occur when
$P_{\rm LZ}=1/2$ and the Stückelberg phase is fully constructive, $\sin^2\Phi=1$. The ridge $P_{\rm LZ}=1/2$ is given by
\begin{equation}
\begin{aligned}
\lambda_1^{\ast}(\delta,\kappa_y)
&=\left(\frac{2\ln2}{\pi}\right)^{\!1/2}\frac{(-\delta-\kappa_y^2)^{1/4}}{(-\delta)^{1/2}}\\
&\approx \left(\frac{2\ln2}{\pi}\right)^{\!1/2}(-\delta)^{-1/4}
\left[1-\frac{\kappa_y^2}{4|\delta|}+O(|\delta|^{-2})\right]
\end{aligned}
\end{equation}
with the last line valid for $|\delta|\gg \kappa_y^2$. The constructive-interference points along this ridge satisfy
\begin{equation}
\zeta\!\left(\delta,\lambda_1^{\ast}\right)
+\varphi_S\!\left(\frac{\ln2}{2\pi}\right)
+\phi_g(\delta,\kappa_y)
=\left(n+\tfrac12\right)\pi.
\end{equation}
For $|\delta|\gg1$ the dynamical phase has the asymptotic form $\zeta(\delta,\lambda_1)=\frac{4}{3}\tau_0^3+O(1)$ so that for $\kappa_y\approx 0$ and large $|\delta|$,
\begin{equation}
\begin{aligned}
\delta_n^\ast &\approx -(3\pi/4)^{2/3} n^{2/3} \approx -1.77\,n^{2/3}\\
\lambda_{1,n}^\ast &\approx \frac{2^{5/6}\sqrt{\ln 2}}{3^{1/6}\pi^{2/3}} n^{-1/6}\approx 0.58 n^{-1/6},
\end{aligned}
\end{equation}
for integers $n\gg 1$.

\subsection{DDP method, $p$-wave, $\delta+\kappa_y^2>0$}
\label{subsec:DDP_pwave}

We now consider the gapped regime $\delta+\kappa_y^2>0$ for the chiral $p$-wave tunneling.
Setting $\kappa_x=0$ (a choice of time origin), the dimensionless two-level Hamiltonian is
\begin{equation}
    H(\tau)
    =(\tau^2+\delta+\kappa_y^2)\sigma^z
     +\lambda_1\big(\tau\sigma^x-\kappa_y\sigma^y\big),
\end{equation}
with instantaneous level splitting
\begin{equation}
    \mathcal{E}(\tau)
    =2\sqrt{(\tau^2+\delta+\kappa_y^2)^2+\lambda_1^2(\tau^2+\kappa_y^2)}.
    \label{eq:Epwave}
\end{equation}
Since $\delta+\kappa_y^2>0$, the adiabatic levels do not cross on the real $\tau$ axis, and the
Landau--Zener picture is not directly applicable. Once again, we employ the Dykhne--Davis--Pechukas
(DDP) method~\cite{Dykhne1962,Davis1976Apr}, which estimates the interband transition amplitude from
complex-time saddle-points $\tau^\ast$ satisfying $\mathcal{E}(\tau^\ast)=0$ in the upper half-plane.
The condition $\mathcal{E}(\tau^\ast)=0$ is equivalent to
\begin{equation}
    (\tau^{\ast 2}+\delta+\kappa_y^2)^2+\lambda_1^2(\tau^{\ast 2}+\kappa_y^2)=0.
\end{equation}
Writing $u=\tau^{\ast 2}$, this becomes a quadratic equation
\begin{equation}
    u^2+(2\delta+2\kappa_y^2+\lambda_1^2)u+\bigl(\delta+\kappa_y^2\bigr)^2+\lambda_1^2\kappa_y^2=0,
\end{equation}
with discriminant that simplifies to
\begin{equation}
    (2\delta+2\kappa_y^2+\lambda_1^2)^2-4\Big[(\delta+\kappa_y^2)^2+\lambda_1^2\kappa_y^2\Big]
    =\lambda_1^2(\lambda_1^2+4\delta).
\end{equation}
Thus $u$ has two real negative roots, and the corresponding complex solutions lie on the
imaginary axis:
\begin{equation}
    \tau^\ast=\pm i a,\qquad \tau^\ast=\pm i b,\qquad 0<a<b,
\end{equation}
where
\begin{equation}
\label{eq:ab_defs}
\begin{aligned}    a^2&=\delta+\kappa_y^2+\frac{\lambda_1^2}{2}-\frac{\lambda_1}{2}\sqrt{\lambda_1^2+4\delta},
\\    b^2&=\delta+\kappa_y^2+\frac{\lambda_1^2}{2}+\frac{\lambda_1}{2}\sqrt{\lambda_1^2+4\delta}.
\end{aligned}
\end{equation}
In the standard, leading DDP approximation, we keep only the dominant contribution coming from the saddle-point closest to the real
axis, $\tau_c=i a$, with other contributions exponentially suppressed. We write
\begin{equation}
    \mathcal{D}_+ =\int_{0}^{\tau_c}\mathcal{E}(\tau)\,d\tau.
\end{equation}
Since $\tau_c$ is purely imaginary, we may integrate along the imaginary axis by setting
$\tau=i y$, $0\le y\le a$, which gives $\mathcal{D}_+= i\int_0^{a}\mathcal{E}(i y)dy$ with
\begin{equation}
    \mathcal{E}(i y)=2\sqrt{(\delta+\kappa_y^2-y^2)^2+\lambda_1^2(\kappa_y^2-y^2)}.
\end{equation}
Using the definitions \eqref{eq:ab_defs}, the square root factorizes as
\begin{equation}
    (\delta+\kappa_y^2-y^2)^2+\lambda_1^2(\kappa_y^2-y^2)=(a^2-y^2)(b^2-y^2),
\end{equation}
so the imaginary part of the action reduces to
\begin{equation}
\label{eq:ImD_integral}
    \Im\mathcal{D}_+
    =\int_0^{a}\mathcal{E}(i y)\,dy
    =2\int_0^{a}dy\;\sqrt{(a^2-y^2)(b^2-y^2)}.
\end{equation}
This integral can be expressed in closed form in terms of complete elliptic integrals of the first
and second kind, $K(m)$ and $E(m)$, with modulus $m=a^2/b^2$:
\begin{equation}
\label{eq:ImD_elliptic}
    \Im\mathcal{D}_+
    =\frac{2b}{3}\Big[(a^2+b^2)\,E(m)-(b^2-a^2)\,K(m)\Big]
\end{equation}
The leading DDP estimate for the asymptotic interband tunneling probability is then
\begin{equation}
\label{eq:DDP_pwave_prob}
    P(\delta,\lambda_1)
    \approx \exp(-2\,\Im\mathcal{D}_+).
\end{equation}
In contrast to the $s$-wave case, where symmetry-related complex crossings produce an interference
factor $\sin^2(\Re\mathcal{D}_+)$, here $P(\delta,\lambda_1)$ has a purely exponential form.

\par For the dominant transverse channel $\kappa_y=0$ and small $\delta$,
we have the simplification
\begin{equation}    a=\frac{\sqrt{\lambda_1^2+4\delta}-\lambda_1}{2}\approx \frac{\delta}{\lambda_1}\qquad
    (\lambda_1\gg\sqrt{\delta}),
\end{equation}
and the action admits the asymptotic expansion
\begin{equation}
\label{eq:ImD_asymp_ky0}
    \Im\mathcal{D}_+
    =\frac{\pi}{2}\frac{\delta^2}{\lambda_1}+O\!\left(\frac{\delta^3}{\lambda_1^3}\right).
\end{equation}
Therefore,
\begin{equation}
\label{eq:P_asymp_ky0}
    P
    \approx \exp\!\left[-\pi\frac{\delta^2}{\lambda_1}\right],
\end{equation}
so at fixed $\delta$ the probability increases monotonically with $\lambda_1$, in contrast with the $s$-wave case, but reminiscent of the LZ problem. 

For fixed nonzero $\kappa_y$, we have $\Im\mathcal{D}_+\sim(\pi/2)\lambda_1\kappa_y^2$
at large $\lambda_1$, implying strong suppression away from $\kappa_y=0$; thus the net transmission is
dominated by a narrow window of small $|\kappa_y|$ in the large-$\lambda_1$ regime.

\section{WKB approach and finite device size}

Our analysis has assumed a device length $L_x$ much greater than other length scales. Here, we comment on the effect of finite $L_x$ and the device sizes necessary to observe our predictions. The essential takeaway is that the device must be large enough that $L_x \gtrsim E_{\mathrm{g}} / eF$, where $E_{\mathrm{g}}$ is the bandgap, in order for our predictions to be clearly observed.    

\par 

To correctly account for finite $L_x$ in temporal gauge, we need to choose $A_x(t) = -Ft\chi_{-L_x/2<x<L_x/2}$, where $\chi$ is an indicator function. Due to the mixed spatial and time dependence, it is far more convenient to work in the static gauge, in which there is a potential $V(x) = eFx\chi_{-L_x/2<x<L_x/2}$. Then it is standard to estimate the tunneling probability using the WKB approximation~\cite{Nandkishore2011Aug,scammell2025zener}, which we work out here. Taking $s$-wave tunneling for simplicity, the wave-function is written as a sum of plane waves with wavevectors $\kappa(x)$ satisfying 
\begin{equation}    \left(\frac{\hbar^2\kappa(x)^2+p_y^2}{2m^*} + \Delta\right)^2 +T_0^2 -(V(x)-E)^2 = 0.
\end{equation}
We adopt $E = 0$ for now, and the result will apply to states centered near $x=0$. The solution for $\kappa$ is 
\begin{equation}
    \kappa(\xi)^2 = k_{\mathrm{Z}}^2\left(\pm \sqrt{\xi^2-\lambda^2} - (\delta +\kappa_y^2)\right)
\end{equation}
where we've defined rescaled position $\xi = x/\ell_{\mathrm{Z}}$. Introduce a shorthand 
\begin{equation}
    B = \delta + \kappa_y^2.
\end{equation}
There are crossing points 
\begin{equation}
    \begin{aligned}
        \xi_1 &= \lambda\\
        \xi_2 &= \sqrt{B^2 + \lambda^2}
    \end{aligned}
\end{equation}
at which the solutions $\kappa(\xi)$ change their real or imaginary character. In the region $|\xi|< \xi_1$, there are four complex solutions: 
\begin{equation}
\kappa(\xi) = \pm k_{\mathrm{Z}}\sqrt{\pm i \sqrt{\lambda^2-\xi^2} - B}.
\end{equation}
Of these, we keep only the solutions with positive imaginary part, which correspond to tunneling from left to right. In the region $\xi_1 < |\xi| < \xi_2$, there are two decaying solutions with different decay constants. We keep only the solution with longer decay length, namely
\begin{equation}
    \kappa(x) = ik_{\mathrm{Z}} \left( B- \sqrt{\xi^2-\lambda^2}\right)^{1/2}.
\end{equation}
Finally, for $|\xi| > \xi_2$, we have two propagating solutions
\begin{equation}
    \kappa(x) = \pm k_{\mathrm{Z}} \left(\sqrt{\xi^2-\lambda^2} - B\right)^{1/2}
\end{equation}

\paragraph*{Interfering WKB amplitude.}
Because the forbidden region consists of an ``inner'' part with two degenerate complex-conjugate trajectories and an ``outer'' part with a single dominant decay channel on each side, the net left-to-right tunneling amplitude for fixed $\kappa_y$ takes the generic form~\cite{Nandkishore2011Aug}
\begin{equation}
    A_{\kappa_y}
    =\Big[c\,e^{-S_1(B,\lambda)}+c^*\,e^{-S_1^*(B,\lambda)}\Big]\,   e^{-2S_2(B,\lambda)},
    \label{eq:Ak}
\end{equation}
where $S_1$ is the (complex) action accumulated across $|\xi|<\xi_1$, and $S_2$ is the (real) action accumulated across $\xi_1<|\xi|<\xi_2$ on one side. 
Explicitly, we have
\begin{align}
    S_1(B,\lambda)
    &=\int_{-\xi_1}^{\xi_1} d\xi\;
    \sqrt{-B+i\sqrt{\lambda^2-\xi^2}},
    \label{eq:S1Zener}\\[4pt]
    S_2(B,\lambda)
    &=\int_{\xi_1}^{\xi_2} d\xi\;
    \Big(B-\sqrt{\xi^2-\lambda^2}\Big)^{1/2}.
    \label{eq:S2Zener}
\end{align}
The complex phase ${\rm Im}\,S_1$ produces oscillations, while ${\rm Re}\,S_1$ and $S_2$ set the overall suppression. Writing $c=|c|e^{i\phi}$, the transmission probability $T_{\kappa_y}=|A_{\kappa_y}|^2$ becomes
\begin{equation}
    T_{\kappa_y}
    =\mathcal N\,
    e^{-\,4S_2(B,\lambda)-2\,{\rm Re}\,S_1(B,\lambda)}
    \cos^2\!\Big({\rm Im}\,S_1(B,\lambda)+\phi\Big),
    \label{eq:TkWKB}
\end{equation}
where $\mathcal N=4|c|^2$,
which has the expected structure of an oscillation modulated by an exponential suppression. The complex prefactor $c$ can, in principle, be obtained by matching across $\xi_1$ and $\xi_2$ (including, in full generality, the subdominant decaying solutions and exponentially growing branches~\cite{scammell2025zener}), but for our purposes it is sufficient to treat $(\mathcal N,\phi)$ as fitting parameters as in Ref.~\cite{Nandkishore2011Aug}. We found that the fit provided by the WKB approach, though reasonable, is inferior to the fit provided by the DDP approach.

\par 

\paragraph*{Finite-$L_x$ criterion.}
The construction above assumes that the turning points at $\pm x_2$ lie inside the field region, i.e.
\begin{equation}
    \xi_L\equiv \frac{L_x}{2\ell_Z} \gtrsim \xi_2=\sqrt{(\delta+\kappa_y^2)^2+\lambda^2}.
    \label{eq:LxCriterion}
\end{equation}
For the dominant channels near $\kappa_y\simeq 0$ and moderate $\lambda$, this reduces parametrically to
\begin{equation}\label{eq:Lxconstraint}
    L_x \gtrsim \frac{2\Delta}{eF} \equiv \frac{E_\mathrm{g}}{eF},
\end{equation}
so that the device contains the full classically forbidden region. When $\xi_L<\xi_2$, this approach breaks down and interlayer charge transfer is highly suppressed.\par 

The condition $L_x \gtrsim E_\mathrm{g}/eF$ places no serious constraints on experimental setups. For example, for $E_\mathrm{g}\simeq 10$--$100~{\rm meV}$, a $L_x=1~\mu{\rm m}$ device requires only $F \gtrsim 0.01$--$0.1~{\rm V}/\mu{\rm m}$ to enter the long-device regime. Much larger average in-plane fields are routinely achieved in nanoscale 2D devices; for instance, monolayer MoS$_2$ transistors with $L_x\simeq 50~{\rm nm}$ operated at $V_{\rm sd}\sim 1~{\rm V}$ correspond to average fields of order $F\sim 20~{\rm V}/\mu{\rm m}$~\cite{daus2021high}, satisfying Eq.~\eqref{eq:Lxconstraint} with a factor of $10^2$ to spare.

\section{Landauer--B\"uttiker}

This section makes precise why a Landauer--B\"uttiker (LB) conductance formula applies to our setup. The standard formula relates current to the transmission integrated over the distribution functions of the leads: 
\begin{equation}
    I = \frac{e}{h} \int_{-\infty}^{\infty} \mathrm{d}\mu \, \left(n_{\mu_L} - n_{\mu_R}\right) \mathbf{T}(\mu)
\end{equation}
where $n_\mu$ is the Fermi-Dirac distribution at chemical potential $\mu$ and $\mathbf{T}(\mu)$ is the net transmission at chemical potential $\mu$, a sum of contributions from all conducting channels at $\mu$. In our case, we have an in-place field $F\hat x$ which, in static gauge, corresponds to a potential bias $V(x) = -eFx$; it is not obvious that this can be interchanged with the chemical potential in the LB formula. Instead, to derive the net current in this scenario, let us consider the theoretically simpler setting of a one-dimensional ring. Let us induce an electric field $F$ around the ring by threading a magnetic flux and calculate the net current. Let the ring have lattice spacing $a$ and total size $L$, and let $\Delta k = 2\pi/a$, and let there be a conduction and valence band with an interlayer tunneling probability $P$ as in our setting. For electrons in the valence band, we have 
\begin{equation}
    \dv{p}{t} = \hbar \dv{k}{t} = eE.
\end{equation}
The time it takes for an electron to traverse the Brillouin zone is then given by $\tau = \hbar \Delta k/eF$, during which time there is a probability $P$ for an electron to hop from valence to conduction band. The rate of change in electron number of the conduction band is then 
\begin{equation}
    \dv{N}{t} = gL \frac{\Delta k}{2\pi} \frac{P}{\tau} 
\end{equation}
where $g$ is an optional degeneracy. Dividing by $L$ and simplifying, the rate of change of electron density in the conduction band is 
\begin{equation}
    \dv{n}{t} = \frac{e}{h} FgP. 
\end{equation}
The induced current density can then be expressed as
\begin{equation}
    \begin{aligned}
        I &= -evn \\
        &= -ev\dv{n}{t} \frac{L}{v}\\
        &=\frac{-e^2}{h} FgP L\\
        &= \frac{e^2}{h}gP V
    \end{aligned}
\end{equation}
where $v$ is velocity, $L/v$ is the time to traverse the sample, and $V = -FL$. This relates $P$ to the conductance as $G = \frac{e^2}{h}gP$, which is the result we use the main text, and which coincides with the Landauer-B\"uttiker formula when $\mathbf{T}(\mu)$ is taken approximately $\mu$-independent and $\mu_R-\mu_L \approx \Delta\mu$ is taken small.\par

\section{Interaction effects}
\label{sec:interactions}

In this section, we describe the effect of many-body interactions on resonant Zener interferometry. We focus on the regime of weak interactions where exciton formation is negligible, and assume that mass renormalization is already captured by the effective single-particle Hamiltonian. The primary remaining concern is the dynamical phase shift induced by the attractive interaction between the tunneling electron and the hole left behind in the valence band. We model this interaction using a Coulomb form:
\begin{equation}
    U_{\mathrm{eh}}(\vec r) = \frac{-e^2}{\varepsilon \sqrt{d^2+r^2}},
\end{equation}
where $\vec r=(x,y)$ is the electron-hole separation, $d$ is the interlayer distance, and $\varepsilon$ is the dielectric constant. To evaluate the accumulated dynamical phase, we first determine the trajectory of the carriers. Applying the Ehrenfest theorem to the single-particle Hamiltonian $H(t)=H_0\!\bigl(p_x+eF t,\;p_y\bigr)$, we find 
\begin{equation}
    \langle x(t)\rangle = \frac{eF}{2m^\ast}\langle \sigma^z\rangle t^2,
\end{equation}
up to a constant shift of $x$ and $t$. Let us focus on the band-overlap regime $\Delta <0$. The dynamical phase $\zeta(\delta,\lambda)$ from the main text receives a correction $\Delta\Phi$ from the interaction $U_{\mathrm{eh}}$ between the electron and hole in the time interval between Landau-Zener avoided crossings. In the notation of the main text, these events occur at times $\pm \tau_0t_{\mathrm{Z}} = \pm t_{\mathrm{Z}}\sqrt{-\delta} $. Hence
\begin{equation}
    \Delta\Phi = \frac{1}{\hbar} \int_{-t_{\mathrm{Z}}\sqrt{-\delta}}^{t_{\mathrm{Z}}\sqrt{-\delta}} U_{\mathrm{eh}}(t) \, \mathrm{d}t \approx \frac{-e^2}{\varepsilon \hbar} \int_{-t_{\mathrm{Z}}\sqrt{-\delta}}^{t_{\mathrm{Z}}\sqrt{-\delta}} \frac{\mathrm{d}t}{\sqrt{d^2 + \left(\frac{eF}{m^\ast}t^2\right)^2}}
\end{equation}
where we've approximated using the expected value of electron-hole separation. This integral can be evaluated explicitly, but the important feature for our purposes is that in the weak-field regime $F\to 0$, $ \Delta\Phi \sim \frac{1}{\sqrt{F}}$. Since the primary Stückelberg interference phase scales as $\Phi_{\mathrm{St.}} \sim 1/F$, the interaction correction becomes negligible in this limit ($F^{-1/2}/F^{-1} \to 0$). At large fields, $\Delta \Phi\sim 1/F$, rendering it negligible compared to the $O(1)$ primary St\"uckelberg phase. Hence, in both regimes, interaction effects will not spoil the salient features of resonant Zener interferometry.

\section{In-plane magnetic field}\label{sec:Bfield}

Here, we explore the effects of an additional in-plane \textit{magnetic} field. Let us consider an in-plane magnetic field $\vec B = B\hat x$, in addition to the in-plane electric field $F\hat x$. In the gauge $\vec A = (-Ft,-Bz,0)$, minimal coupling shifts the transverse momentum $p_y \to p_y-eBz$. For a bilayer with layer separation $d$ (layers at $z=\pm d/2$), the $s$-wave tunneling Hamiltonian reads
\begin{equation}
H =
\begin{pmatrix}
\frac{(p_x-eFt)^2}{2m^*}+\frac{(p_y+edB/2)^2}{2m^*}+\Delta & T_0\\
T_0 & -\frac{(p_x-eFt)^2}{2m^*}-\frac{(p_y-edB/2)^2}{2m^*}-\Delta
\end{pmatrix}.
\end{equation}
Introducing Zener scales $\ell_\mathrm{Z}=(\hbar^2/2m^*eF)^{1/3}$, $k_\mathrm{Z}=\ell_\mathrm{Z}^{-1}$, $E_\mathrm{Z}=eF\ell_\mathrm{Z}$, and dimensionless variables
$\tau=(p_x-eFt)/\hbar k_\mathrm{Z}$, $\kappa_y=p_y/\hbar k_\mathrm{Z}$, $\delta=\Delta/E_\mathrm{Z}$, $\lambda=T_0/E_\mathrm{Z}$, and
$b=(edB/2\hbar k_\mathrm{Z})$, we obtain a rescaled Hamiltonian
\begin{equation}
H(\tau)=
\begin{pmatrix}
\tau^2+(\kappa_y+b)^2+\delta & \lambda\\
\lambda & -\tau^2-(\kappa_y-b)^2-\delta
\end{pmatrix}.
\end{equation}
The trace of $H(\tau)$ is $4\kappa_y b$ and contributes only a global phase. Subtracting $(2\kappa_y b)\mathbf{1}$ yields the
trace-free two-level problem
\begin{equation}
H'(\tau)=\bigl(\tau^2+\delta+\kappa_y^2+b^2\bigr)\sigma_z+\lambda\sigma_x,
\end{equation}
so the interlayer tunneling probability depends only on the effective bandgap
$\delta_{\rm eff}=\delta+\kappa_y^2+b^2$.
Therefore, the conductance in the presence of $B$ is obtained from the $B=0$ result by the replacement
$\delta\to\delta+b^2$:
\begin{equation}
G = \frac{ge^2}{h}\frac{L_y k_Z}{2\pi}\int \mathrm{d}\kappa_y\,P(\delta+b^2+\kappa_y^2,\lambda).
\end{equation}
Hence the effect of in-plane field is captured implicitly by Fig. 2 of the main text, with $x$-axis $\Delta/E_{\mathrm{Z}}$ replaced by $\Delta/E_{\mathrm{Z}} + (edB/2\hbar k_{\mathrm{Z}})^2$. With sufficient resolution, this may also provide an independent method of determining the precise interlayer separation $d$. In the band-overlap regime ($\delta_{\rm eff}<0$), $P$ exhibits Landau--Zener--St\"uckelberg oscillations as a function of
$\delta_{\rm eff}$; at fixed $\Delta$ and $F$, sweeping $B$ changes $\delta_{\rm eff}$ through $b^2\propto B^2$, producing
interferometric oscillations in $G(B)$. An in-plane magnetic field along $\vec B = B\hat y$, corresponding to a gauge $\vec A = (-Ft +Bz,0,0)$, and indeed an in-plane magnetic field pointing along any direction, would lead to the same conclusions following the argument above with minor modifications.

\bibliography{bib}